\documentclass[a4paper,11pt]{article}
\usepackage{pos}
\usepackage{natbib}
\usepackage{caption}
\usepackage{graphicx} 
\usepackage{multicol} 
\usepackage{array} 
\usepackage{amsmath}
\DeclareMathAlphabet{\mathpzc}{OT1}{pzc}{m}{it}

\title{Pion, Kaon and nucleon gravitational form factors}
\ShortTitle{Pion, Kaon and nucleon GFFs}

\author[a,b]{Z.-Q. Yao}
\author[a,b]{Y.-Z. Xu}
\author[c]{D. Binosi}
\author[d,e]{Z.-F. Cui}
\author[d,e,f]{M. Ding}
\author[a]{K. Raya}
\author[d,e]{\\C. D. Roberts}
\author*[a]{J. Rodríguez-Quintero}

\affiliation[a]{Dpto. Ciencias Integradas, Centro de Estudios Avanzados en Fis., Mat. y Comp., Fac. Ciencias Experimentales, Universidad de Huelva, Huelva E-21071, Spain}

\affiliation[b]{Dpto. Sistemas Físicos, Químicos y Naturales, Univ. Pablo de Olavide, E-41013 Sevilla, Spain}

\affiliation[c]{European Centre for Theoretical Studies in Nuclear Physics and Related Areas, Villa Tambosi, Strada delle Tabarelle 286, I-38123 Villazzano (TN), Italy}

\affiliation[d]{School of Physics, Nanjing University, Nanjing, Jiangsu 210093, China}

\affiliation[e]{Institute for Nonperturbative Physics, Nanjing University, Nanjing, Jiangsu 210093, China}

\affiliation[f]{Helmholtz-Zentrum Dresden-Rossendorf, Bautzner Landstraße 400, D-01328 Dresden, Germany}

\emailAdd{jose.rodriguez@dfaie.uhu.es}

\abstract{
\vspace*{-10ex}
\rightline{\sf NJU-INP 098/25}

\vspace*{6ex}
A unified set of predictions for pion, kaon and nucleon gravitational form factors is obtained using a symmetry-preserving truncation of each relevant quantum field equation. A crucial aspect of the study is the self-consistent characterization of the dressed quark-graviton vertices, applied when probing each quark flavor inside mesons or nucleons. The calculations reveal that each hadron’s mass radius is smaller than its charge radius, matching available empirical inferences; moreover, core pressures are significantly greater than those in neutron stars. This set of predictions is expected to be instrumental as forthcoming experiments provide opportunities for validation.}

\FullConference{The XVIth Quark Confinement and the Hadron Spectrum Conference (QCHSC24)\\
 19-24 August, 2024\\
 Cairns Convention Centre, Cairns, Queensland, Australia\\}

\begin{document}
\maketitle

\section{Introduction}
Nucleons play a central role in the pursuit of understanding matter and its interactions. The properties of these hadrons, which form the heart of atomic nuclei, are deeply connected to the emergent phenomena of quantum chromodynamics (QCD) \cite{Ding:2022ows}. On the other side of the coin, pions and kaons are Nature's most fundamental Nambu-Goldstone bosons, and their existence and properties are intrinsically linked to the emergent dynamics of QCD (\emph{e.g.}, see Ref.\,\cite{Raya:2024ejx}). Gravitational form factors (GFFs) provide a unique window into key hadronic properties, revealing how mass, spin, and internal forces are distributed within the hadron \cite{Polyakov:2018zvc}. The characterization of GFFs is a cornerstone of the scientific programs at modern facilities. Herein, we report a systematic analysis using continuum Schwinger methods (CSMs) \cite{Xu:2023izo,Yao:2024ixu}.

\section{Gravitational Currents}
\label{eq:currents}

Let us consider a pseudoscalar meson $\textbf{P}$, with $f,\bar{h}$ valence-quark and antiquark flavors and mass $m_\textbf{P}$. The expectation value within $\textbf{P}$ of the energy-momentum tensor takes the form \cite{Polyakov:2018zvc}:
\begin{equation}
    \label{eq:EMTPS}
    \Lambda_{\mu\nu}^{\textbf{P}}(K,Q) = 2K_\mu K_\nu A^\textbf{P}(Q^2) + \frac{1}{2}[Q_\mu Q_\nu-Q^2\delta_{\mu\nu}]D^\textbf{P}(Q^2)+2m_\textbf{P}^2 \delta_{\mu\nu}\bar{c}^\textbf{P}(Q^2)\,.
\end{equation}
Here, $Q^2=(p_f-p_i)^2$ denotes the momentum transfer of the external $J^{PC}=2^{++}$ probe (graviton like),  $K=(p_f+p_i)/2$, and $p_{i,f}$ are the momenta of the incoming/outgoing pseudoscalars.  
$A^{\textbf{P}}(Q^2)$, $D^{\textbf{P}}(Q^2)$ and $\bar{c}^{\textbf{P}}(Q^2)$ are the associated GFFs.
A $\theta_2^\textbf{P}(Q^2)=A^\textbf{P}(Q^2)$ and $\theta_1^\textbf{P}(Q^2)=-D^\textbf{P}(Q^2)$ convention is also commonly employed. 
A sum over parton species, ${\mathscr p}$, may be considered implicit in Eq.\,\eqref{eq:EMTPS}. Although scale-dependent when considered independently for each species, the total form factors are naturally scale-invariant. 
The following relations arise from symmetries:
\begin{equation}
\label{eq:symsPion}
    A^{\textbf{P}}(0) = 1\,,\,D^{\textbf{P}}(0)\overset{m_{\textbf{P}}=0}{=}-1\,,\,\bar{c}^{\textbf{P}}(Q^2)=0\,.
\end{equation}
The first identity is a statement of mass normalization; the second stems from the soft-pion theorem \cite{Polyakov:1999gs,Mezrag:2014jka}; and the third is a consequence of energy-momentum conservation.

The nucleon ($\textbf{N}$), or any spin-$1/2$ hadron in general, is characterized by 3 nonzero GFFs, defined from the following gravitational current\,\cite{Polyakov:2018zvc}:
\begin{equation}
    \label{eq:EMTN}
    m_{\textbf{N}}\Lambda_{\mu\nu}^{\textbf{N}}(K,Q) = -\Lambda_+(p_f)[K_\mu K_\nu A^\textbf{N}(Q^2)
    + i K_{\{\mu}\sigma_{\nu\}\rho} Q_\rho J^{\textbf{P}}(Q^2) + \frac{1}{4}(Q_\mu Q_\nu-Q^2\delta_{\mu\nu})D^\textbf{N}]\Lambda_+(p_i)\,,
\end{equation}
where $a_{\{\mu}b_{\nu\}}=(a_\mu b_\nu+a_\nu b_\mu)/2$, and $\Lambda_+$ denotes the projection operator that delivers a positive energy nucleon. 
In principle, a $\bar{c}^{\textbf{N}}(Q^2)$ GFF is defined in analogy to the pseudoscalar case, Eq.\,\eqref{eq:EMTPS}, and it also vanishes by current conservation. 

As with pseudoscalar mesons, $A^\textbf{N}(Q^2)$ and $D^\textbf{N}(Q^2)$ are associated with the bound-state mass and pressure distributions, respectively, while $J^\textbf{N}(Q^2)$ relates to the nucleon spin distribution. Symmetry principles entail $A^\textbf{N}(0)=1$ and $J^\textbf{N}(0)=1/2$ \cite{Polyakov:2002yz}. In contrast, although $D(0)$ is also a conserved charge, its value is determined by dynamics \cite{Polyakov:2018zvc}: \emph{``the last unknown global property of the nucleon''}. 

We describe below the symmetry-preserving evaluation for the GFFs of the pion, kaon and nucleon within a systematically-improvable continuum framework. In this approach, the form factors are naturally calculated at a rigorously defined hadron scale, whereat all hadron properties are entirely expressed by their fully-dressed valence constituents \cite{Cui:2020tdf}.

\section{Symmetry-preserving evaluation of GFFs}
\label{eq:ingredients}

\begin{figure}[t]
    \centering
    \begin{tabular}{cc}
        \includegraphics[width=0.27\textwidth]{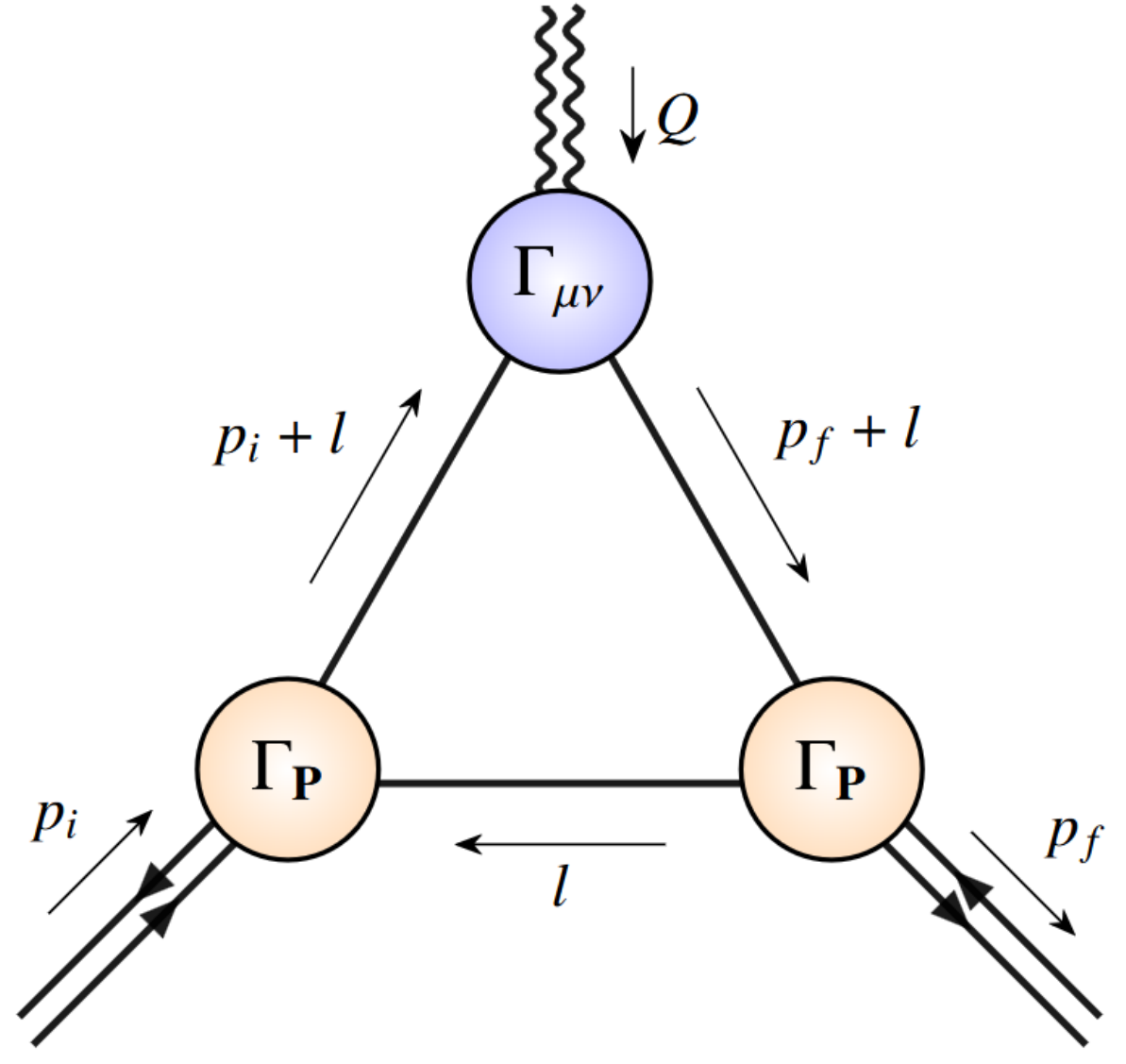} &  
        \includegraphics[width=0.27\textwidth]{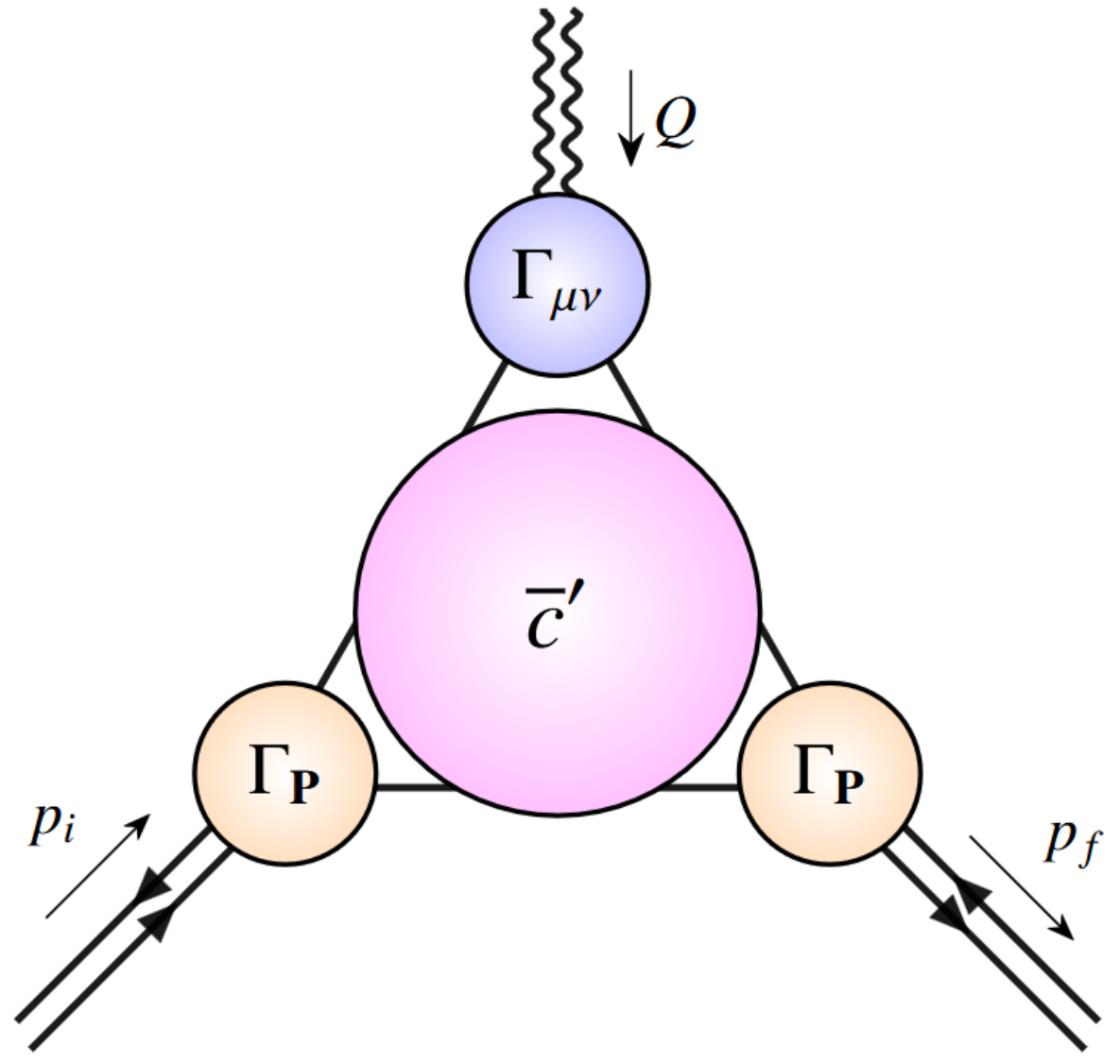} 
    \end{tabular}
    \caption{Probe+meson interaction in RL truncation. \emph{Both panels.} Solid lines -- dressed quark propagators; orange shaded circles -- meson BSAs; blue shaded circles -- dressed probe+quark vertex, $\Gamma_{\mu\nu}$. \emph{Right panel.} Shaded $\mathbb{C}$ region -- gluon-binding contributions to the probe+pion interaction, ensuring $\bar{c}^\textbf{P}(Q^2)\equiv 0$.}
    \label{fig:PsDiagrams}
\end{figure}

The interaction of a pseudoscalar meson with a graviton is sketched in Fig.\,\ref{fig:PsDiagrams}. 
The right diagram plays a role analogous to Ref.\,\cite[Fig.\,3B$^\prime$]{Ding:2019lwe}, which restores momentum conservation in calculations of pion DFs.  
Notwithstanding that, it is not necessary to develop an explicit expression for this term because it only affects $\bar c^\pi(Q^2)$, which is identically zero.
$A^\textbf{P}$ and $D^\textbf{P}$ decouple from the longitudinal projections $Q_\mu \Lambda^\textbf{P}_{\mu\nu}(P,Q)$, $Q_\nu \Lambda^\textbf{P}_{\mu\nu}(P,Q)$.  
The image in the left panel represents the following vertex:
\begin{equation}
    \label{eq:IApi}
    \Lambda_{\mu\nu}^{\textbf{P}f}= N_c \text{tr}_D\int_l^{\Lambda_{\rm R}} \!\! \Gamma_{\mu\nu}^f(l+p_f,l+p_i) S_f(l+p_i)\Gamma^\textbf{P}(l+p_i/2;p_i)S_{\bar{h}}(l) \bar{\Gamma}^\textbf{P}(l+p_f/2;-p_f)S_f(l+p_f)\,,
\end{equation}
where $N_c=3$, $\text{tr}_D$ indicates the trace over Dirac indices, and $\int_l^{\Lambda_{\rm R}}$ denotes a Poincaré-invariant regularization of the four-momentum integral with regulator $\Lambda_{\rm R}$. Further, $S_{f,\bar{h}}$ are the dressed quark/antiquark propagators; $\Gamma^\textbf{P}$, the meson's Bethe-Salpeter amplitude (BSA); and $\Gamma_{\mu\nu}^f$ is the corresponding graviton+quark vertex (QGV). Quark propagators and meson BSAs are evaluated in the rainbow-ladder (RL) truncation \cite{Bender:1996bb, Munczek:1994zz}, following standard procedures \cite{Qin:2020rad, Eichmann:2016yit}.
The construction of the QGV is detailed in Refs.\,\cite{Xu:2023izo, Yao:2024ixu} and sketched below.

The symmetry-preserving diagrams contributing to the  nucleon GFFs are shown in Ref.\,\cite[Fig.\,S.6]{Yao:2024ixu}, extending the electromagnetic current formulation from Ref.\,\cite{Eichmann:2011vu} to the interaction with a gravitational probe. In addition to the dressed quark propagators and QGV, the nucleon Faddeev amplitudes are required.  They are calculated using the RL truncation\,\cite{Eichmann:2016yit, Qin:2019hgk}.

\begin{figure}[t]
    \centering
    \begin{tabular}{cc}
        \includegraphics[width=0.42\textwidth]{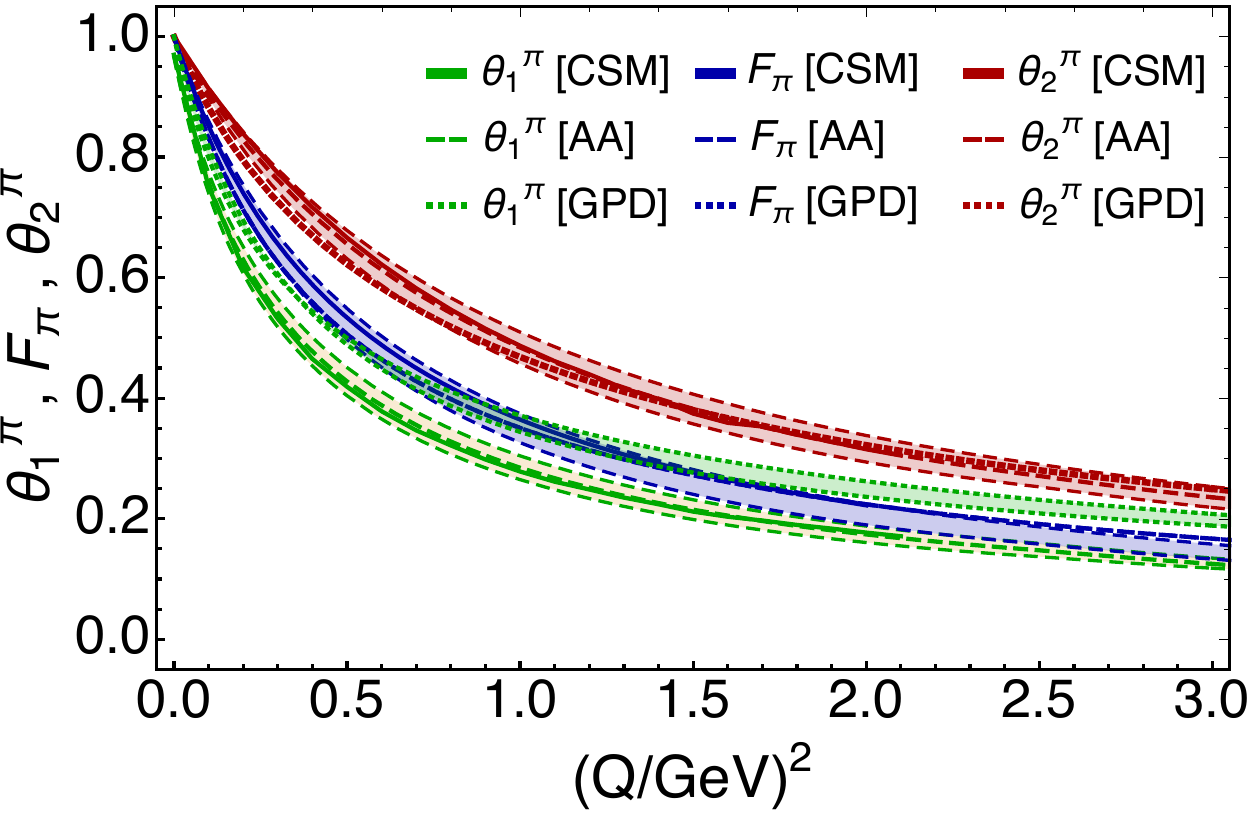} &  
        \includegraphics[width=0.42\textwidth]{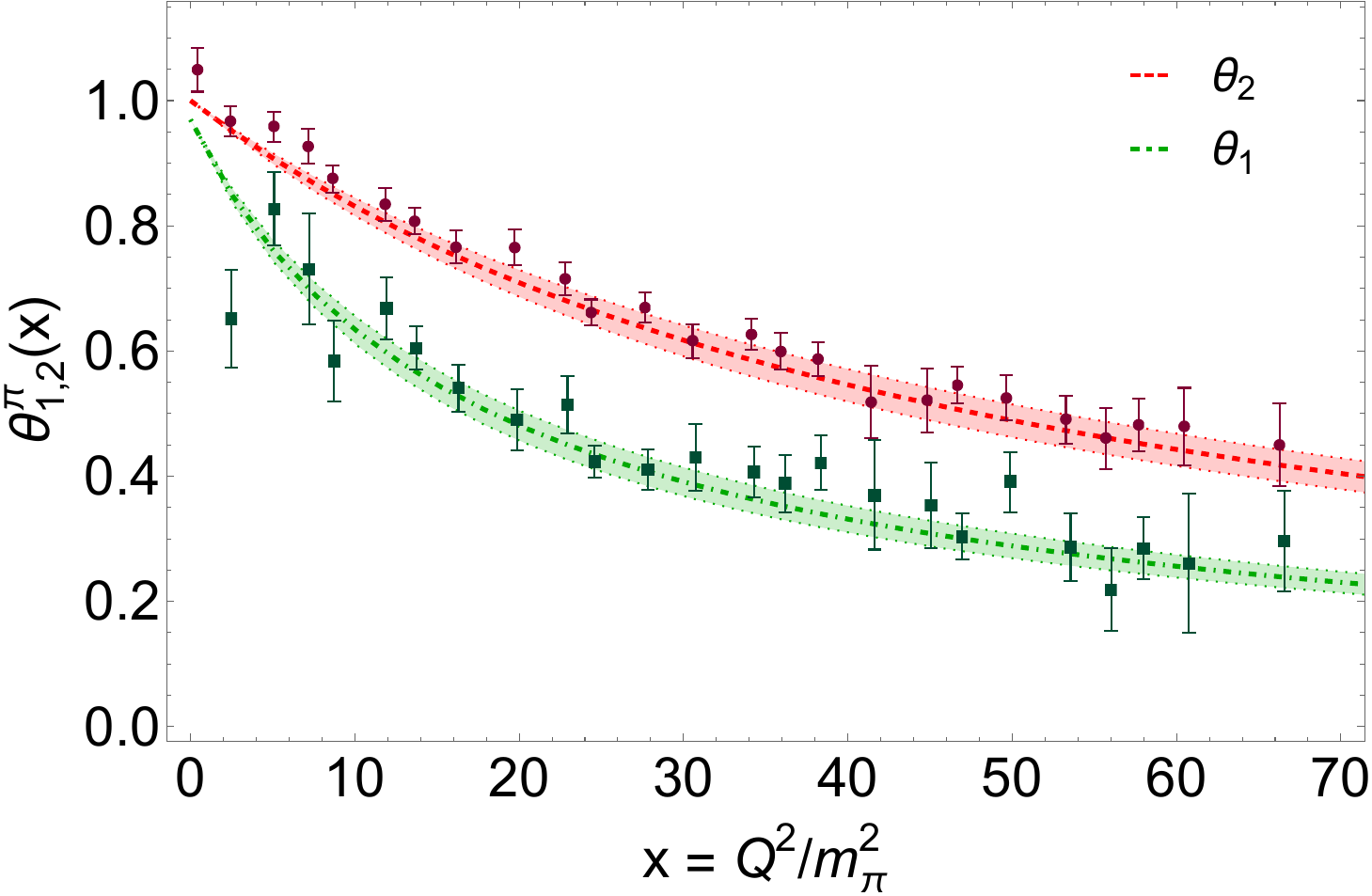} 
    \end{tabular}
    \caption{\emph{Left}. Pion electromagnetic ($F_\pi(Q^2)$) and gravitational form factors. CSM evaluation described herein and in Ref.\,\cite{Xu:2023izo}. AA and  GPD denote results from an algebraic Ansatz \cite{Xu:2023izo}, and a generalized parton distribution approach \cite{Raya:2021zrz}. \emph{Right}. Pion GFFs: CSM and  lattice QCD results ($m_\pi^{lat}=0.17$ GeV)\,\cite{Hackett:2023nkr}.}
    \label{fig:PionGFFs}
\end{figure} 

\begin{figure}[t]
    \centering
    \begin{tabular}{cc}
        \includegraphics[width=0.42\textwidth]{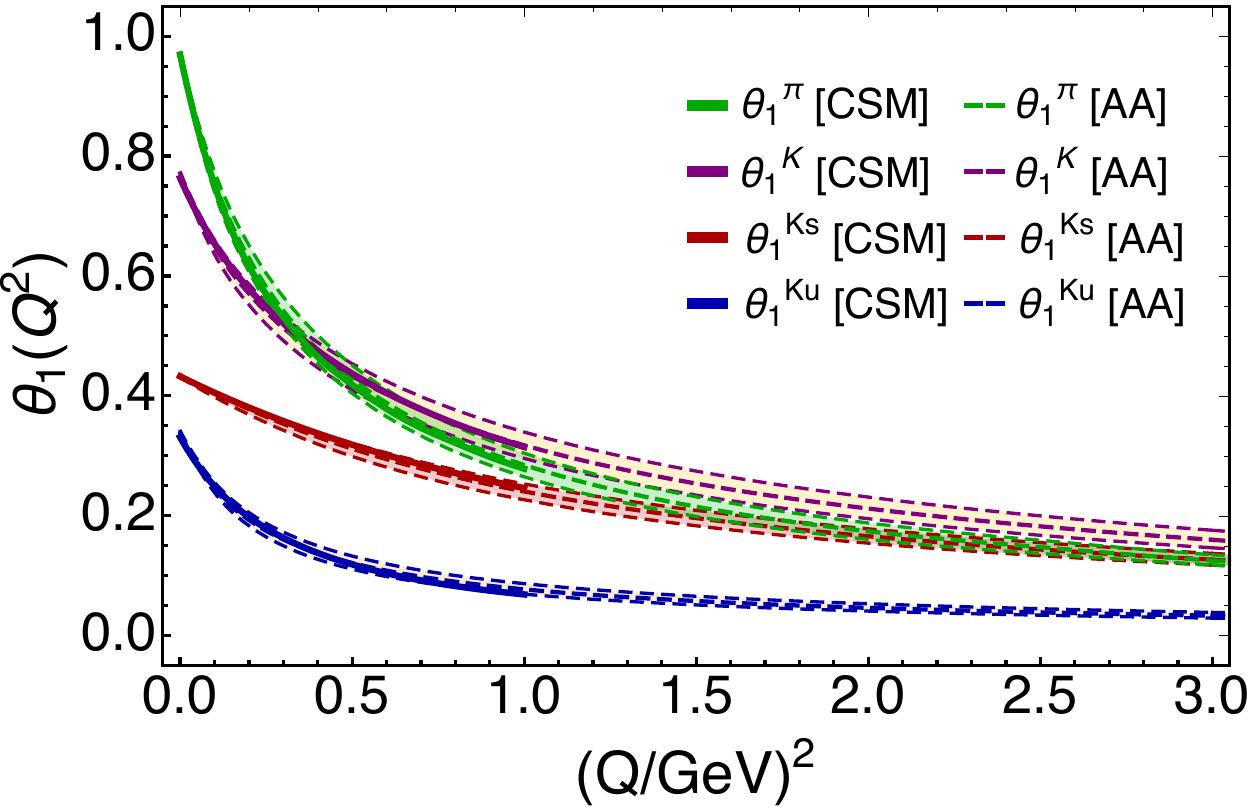} &  
        \includegraphics[width=0.42\textwidth]{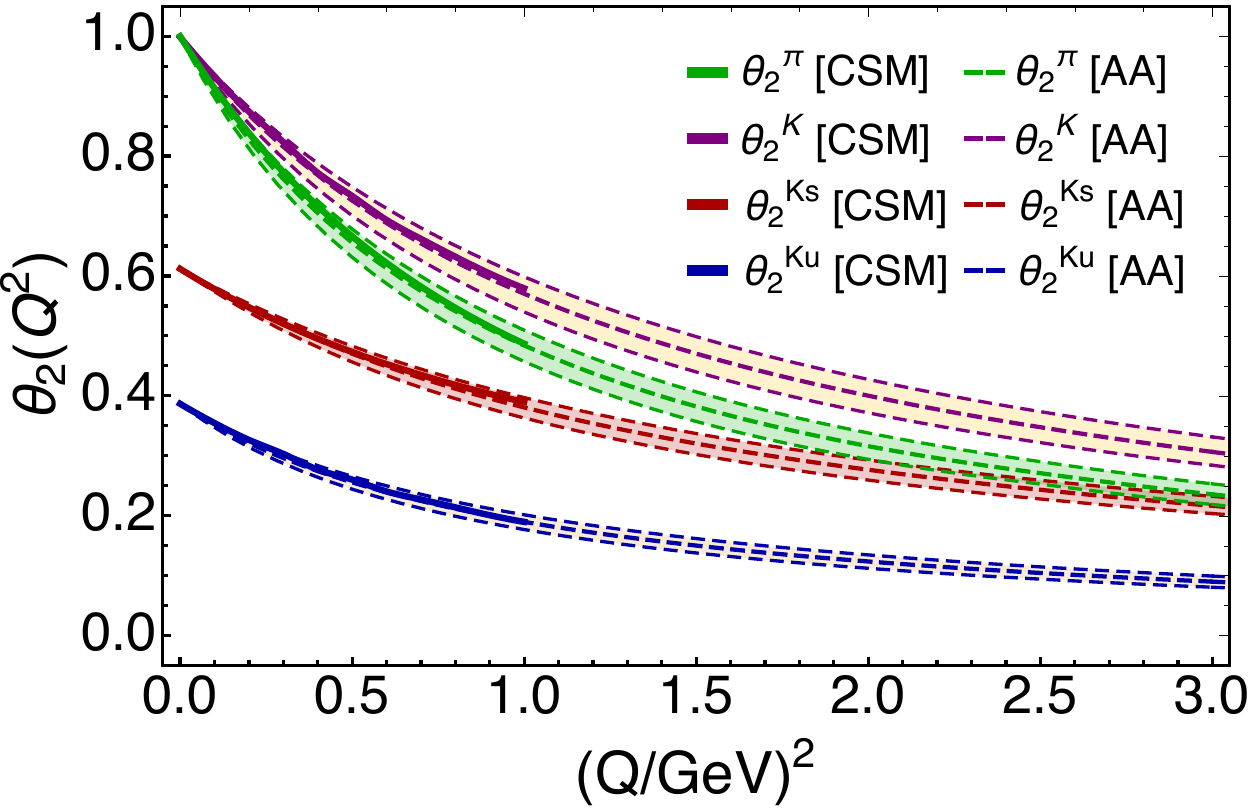} 
    \end{tabular}
    \caption{Pion and (flavor-separated) Kaon GFFs. Legend as in Fig.\,\ref{fig:PionGFFs}.}
    \label{fig:KaonGFFs}
\end{figure}

Analogously to the quark-photon vertex, the QGV obeys a Ward-Green-Takahashi identity (WGTI) \cite{Brout:1966oea}:
\begin{equation}
    \label{eq:WGTI}
    Q_\mu i \Gamma_{\mu\nu}(k,Q)=S^{-1}(k_+)k_{-\nu}-S^{-1}(k_-)k_{+\nu}\,.
\end{equation}
We have defined $k_\pm=k\pm Q/2$ and omitted the flavor indices for simplicity. In RL truncation, this vertex satisfies the following inhomogeneous Bethe-Salpeter equation\,\cite{Xu:2023izo}:
\begin{equation}
\label{eq:BSEQGV}
    i \Gamma_{\mu\nu}(k_+,k_-)=Z_2[i\gamma_\mu k_\nu - \delta_{\mu\nu}(i \gamma \cdot k+Z_m^0 m^\zeta)] + Z_2^2\int_l^\Lambda\mathcal{K}(k-l)[S(l_+)i \Gamma_{\mu\nu}(l_+,l_-)S(l_-)]\,,
\end{equation}
where $m^\zeta$ is the current quark mass, $\zeta$ is the renormalization scale; $Z_m^0$ and $Z_2$ are renormalization constants; and $\mathcal{K}$ is the RL Bethe-Salpeter kernel. 

The general solution of Eq.\,\eqref{eq:BSEQGV}, with complete generality,  may be cast as:
\begin{equation}
    \label{eq:solQGV1}
    \Gamma_{\mu\nu}(k,Q)=\Gamma_{\mu\nu}^{gM}(k,Q)+\Gamma_{\mu\nu}^{gT}(k,Q)\,,
\end{equation}
such that a minimal Ansatz for $\Gamma_{\mu\nu}^{gM}(k,Q)$ reads as follows:
\begin{equation}
\label{eq:solQGV2}
i\Gamma_{\mu\nu}^{g_M}(k,Q)   = i\Gamma_\mu^{\rm BC}(k,Q) k_\nu
-\tfrac{1}{2}\delta_{\mu\nu}[ S^{-1}(k_+) + S^{-1}(k_-)]   +  i T_{\mu\alpha}(Q)T_{\nu\beta}(Q)4 \hat{\Gamma}^2_{\alpha\beta}(k_+,k_-)\,,
\end{equation}
with $T_{\mu\nu}(Q) = \delta_{\mu\nu}-Q_\mu Q_\nu/Q^2$. 
Given a Ball-Chiu-like structure for $\Gamma_\mu^{\text{BC}}(k,Q)$ \cite{Ball:1980ay}, the first two terms guarantee the preservation of the WGTI. The last term in $\Gamma_{\mu\nu}^{g_M}(k,Q)$ emerges from the inhomogeneity:
\begin{equation}
    \Gamma_{0\mu\nu}^2(k,Q)=T_{\mu\alpha}(Q)T_{\nu\beta}(Q)\frac{1}{2}(\gamma_\alpha k_\beta + \gamma_\beta k_\alpha)\,.
\end{equation}
We have defined $\hat{\Gamma}(k,Q)=\Gamma(k,Q)-\Gamma(k,0)$ to ensure the absence of kinematic singularities. Dynamical singularities appear, as expected \cite{Raman:1971jg}, at the pole position of each $I=0$ tensor meson. The final piece in Eq.\,\eqref{eq:solQGV1}, $\Gamma_{\mu\nu}^{gT}(k,Q)$, incorporates scalar meson resonances \cite{Raman:1971jg, Xing:2023eed} via:
\begin{equation}
 \Gamma_{\mu\nu}^{gT}(k,Q) =  T_{\mu\nu}(Q) \Gamma_{\mathbb I}(k;Q)\,;
 \label{eq:solQGV3}
\end{equation}
where $\Gamma_{\mathbb I}(k;Q)$ contains four independent Dirac matrix valued structures\,\cite{Krassnigg:2009zh}, and its determination is described in Ref.\,\cite{Xu:2023izo}. Thus, in total, $\Gamma_{\mu\nu}(k,Q)$ comprises 18 structures: 14 arising from $\Gamma_{\mu\nu}^{gM}(k,Q)$, Eq.\,\eqref{eq:solQGV2}, and the additional 4 from $\Gamma_{\mu\nu}^{gT}(k,Q)$.

\begin{figure}[t]
    \centering
    \begin{tabular}{cc}
        \includegraphics[width=0.45\textwidth]{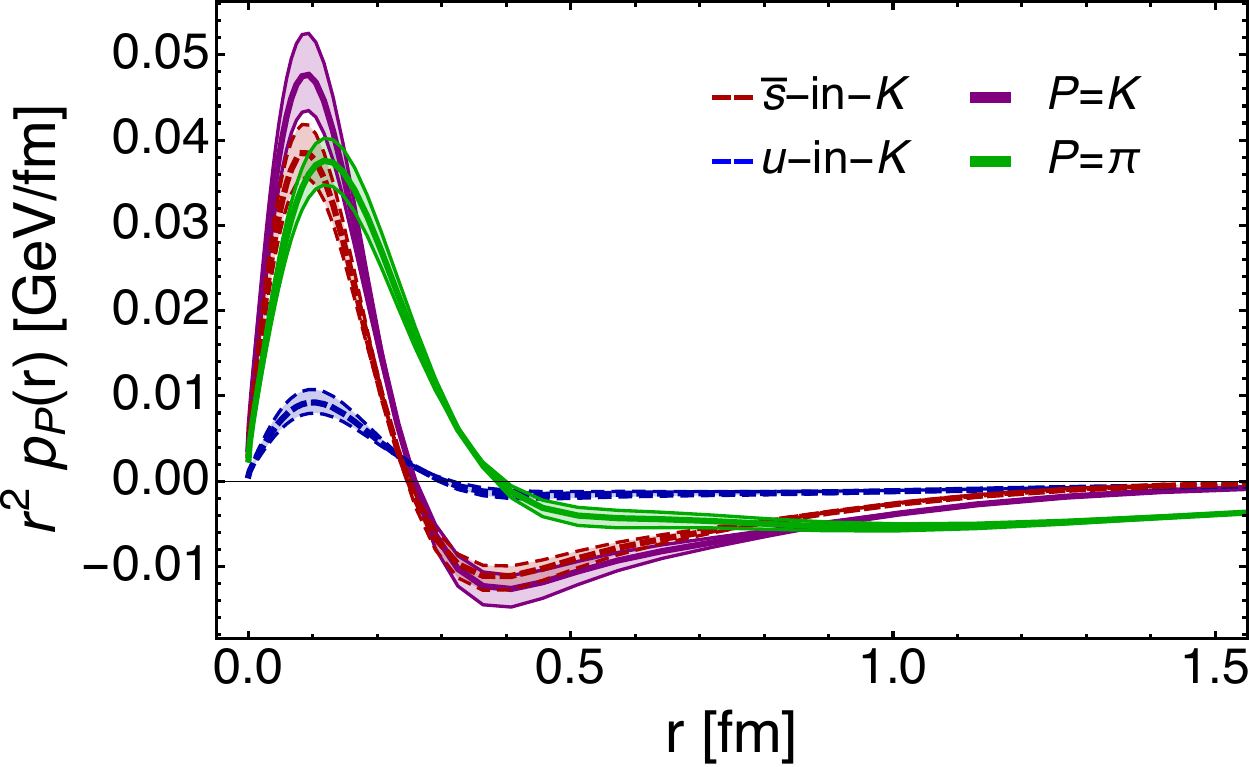} &  
        \includegraphics[width=0.45\textwidth]{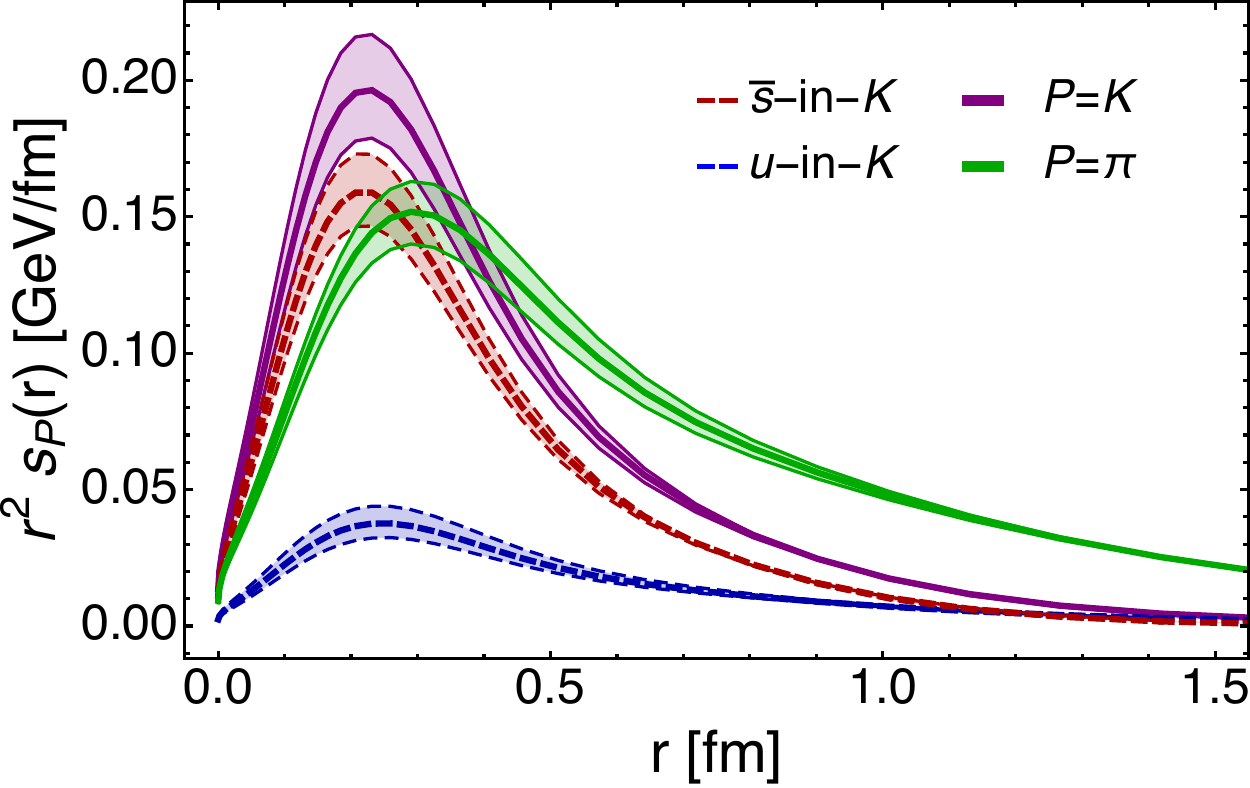} 
    \end{tabular}
    \caption{Pion and (flavor-separated) Kaon pressure and shear forces.}
    \label{fig:PressPS}
\end{figure}

\section{Gravitational form factors}
The predicted pion GFFs are shown in Fig.\,\ref{fig:PionGFFs}. The comparison in the left panel shows that $\theta_2(Q^2)$ exhibits a less pronounced falloff than the corresponding electromagnetic form factor, $F_\pi(Q^2)$; and this, in turn, falls less rapidly than $\theta_1^\pi(Q^2)$. In the right panel, our CSM GFFs are compared with recent lattice QCD results \cite{Hackett:2023nkr}. As an illustration of the systematic impact of the unrealistic lattice pion mass ($m_\pi \approx 0.170$ GeV) on pion observables, compatibility is only revealed when the results appear displayed in terms of $x=Q^2/m_\pi^2$. 

The GFFs of the pion and kaon are compared in Fig.\,\ref{fig:KaonGFFs}, reiterating the ordering of the form factors, and showing that those corresponding to the kaon are slightly harder. We now consider the associated radii:
    $r_{\mathcal{F}}^2=-6(d/dQ^2) \ln \mathcal{F}(Q^2)|_{Q^2=0}\,.$
In the pion (kaon) case:
\begin{equation}
    r_{\theta_1}^\textbf{P}=0.81 \,\text{fm}\,(0.63 \,\text{fm})> r_{F}^\textbf{P}=0.64 \,\text{fm}\,(0.58 \,\text{fm})> r_{\theta_2}^\textbf{P}=0.47 \,\text{fm}\,(0.40 \,\text{fm})\,.
\end{equation}
The average of the different radii indicates that the kaon is more spatially compressed than the pion by approximately a $15\%$. Moreover, we find the mass-to-charge radii ratios to be on the order of $r_{\theta_2}^\pi/r_F^{\pi}\approx 0.74$ and $r_{\theta_2}^K/r_F^K\approx 0.69$. The pion result aligns with a recent empirical determination, $r_{\theta_2}^\pi/r_F^{\pi}\approx 0.79(3)$ \cite{Xu:2023bwv}, and lies within the ballpark of other estimates, \emph{e.g.}, \cite{Raya:2024glv,Wang:2024fjt}.

As far as $\theta_1(Q^2)$ is concerned, we find $\theta_1^\pi(0) = 0.97$ and $\theta_1^K(0)=0.77$, in accord with estimates basaed on chiral effective field theory \cite{Polyakov:2018zvc}. Related pressure and shear force distributions may be defined as follows \cite{Polyakov:2018zvc}:
{\allowdisplaybreaks
\begin{subequations}
\label{EqPressure}
\begin{align}
p_\textbf{P}^f(r)  & =
 \frac{1}{6\pi^2 r} \int_0^\infty d\Delta \,\frac{\Delta}{2 E(\Delta)} \, \sin(\Delta r) [\Delta^2\theta_1^{{\mathscr P}_q}(\Delta^2)] \,, \label{eq:PressureA}\\
 s_\textbf{P}^f (r)  & =
\frac{3}{8 \pi^2} \int_0^\infty d\Delta \,\frac{\Delta^2}{2 E(\Delta)} \, {\mathpzc j}_2(\Delta r) \, [\Delta^2\theta_1^{{\mathscr P}_q}(\Delta^2)] \,, \label{eq:PressureB}
\end{align}
\end{subequations}
where $\Delta = \surd Q^2$,
$E(\Delta)^2 = m_\textbf{P}^2 +\Delta^2/4$
and ${\mathpzc j}_2(z)$ is a spherical Bessel function. The corresponding results are shown in Fig.\,\ref{fig:PressPS}. 
As apparent from the left panel, the meson pressures are positive on $r\simeq 0$ and found there to be significantly greater than those for neutron stars \cite{Ozel:2016oaf}).
On this domain, the meson's dressed-valence constituents are pushing away from each other. 
With increasing separation, the pressure switches sign, indicating a transition to the domain whereupon confinement forces exert their influence on the pair. The zeros occur at the following locations (in fm):
$r_c^\pi=0.39(1)$, $r_c^K=0.26(1)$, $r_c^{K_{u}}=0.30(1)$, $r_c^{K_{\bar s}}=0.25(1)$. The shear pressures are drawn in the right-panel of Fig.\,\ref{fig:PressPS}. Evidently, they are maximal in the neighborhood of $r_c^\textbf{P}$, whereat the forces driving the quark and antiquark apart are being overwhelmed by attractive confinement pressure. 

The nucleon GFFs are displayed in Fig.\,\ref{fig:NucleonGFFs}. The symmetry-preserving character of the CSM analysis is evident in the values of $A(0), J(0)$. Also of interest is the agreement with recent lattice QCD results \cite{Hackett:2023rif}, both in the total form factors and in the parton species separations. 
The species separations were obtained using the all-orders scheme \cite{Cui:2020tdf, Yin:2023dbw} to evolve the form factors calculated at the hadron scale up to the one at which they are estimated from lattice QCD.  
As discussed in Ref.\,\cite{Yao:2024ixu}, the mass radius is smaller than the charge radius: $r_{\text{mass}}/r_{\text{charge}}\approx 0.64(4)$. 
The nucleon pressure and shear force distributions were also calculated \cite[Fig.\,3]{Yao:2024ixu}).  The nucleon pressure is found to possess a near-core value that is approximately half as large as that in the pion. 
The CSM prediction for the nucleon $D$-term is $D(0) = - 3.114(10)$, a result that is consistent with the data-informed extraction in Ref.\,\cite{Cao:2024zlf}: $D(0) = -3.38_{-0.32}^{+0.26}$.
A discussion of the trace anomaly GFF may be found elsewhere in these proceedings \cite{Binosi:2025kpz}.

\begin{figure}[t]
    \centering
    \begin{tabular}{lll}
        \hspace{-2ex}\includegraphics[width=0.325\textwidth]{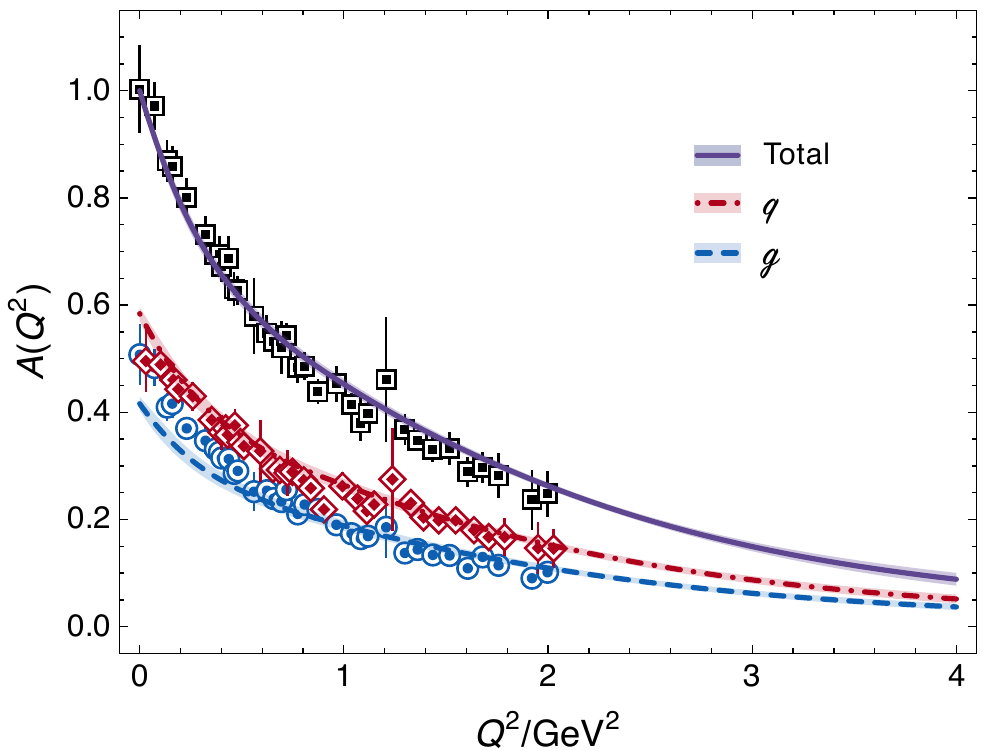} &  
        \hspace{-2ex}\includegraphics[width=0.325\textwidth]{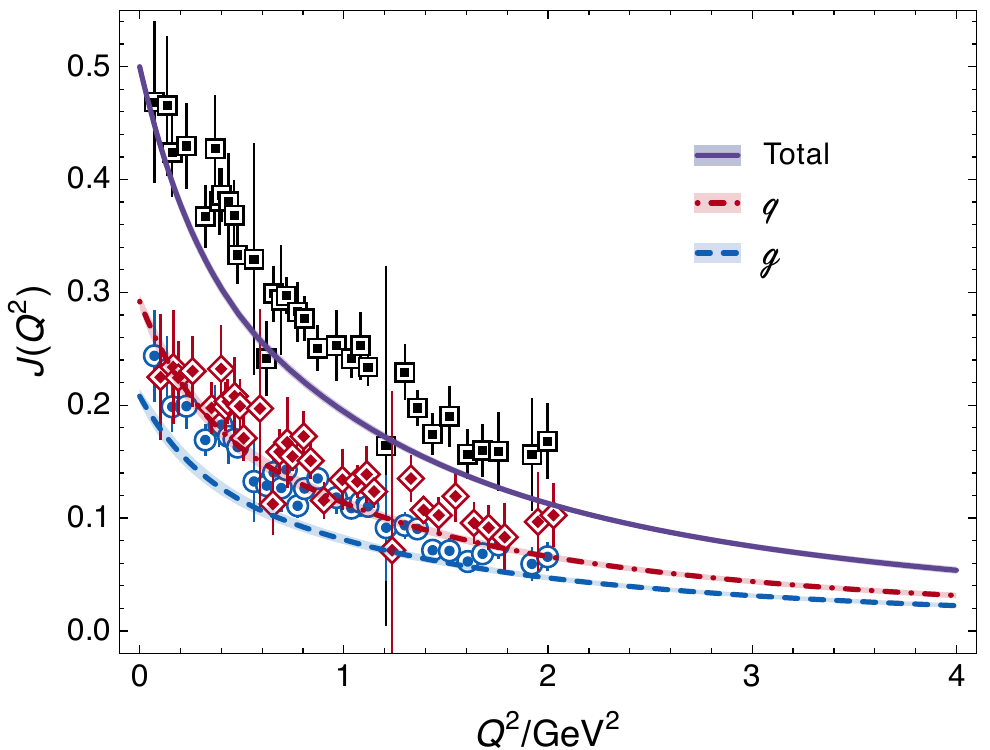} &  
        \hspace{-2ex}\includegraphics[width=0.325\textwidth]{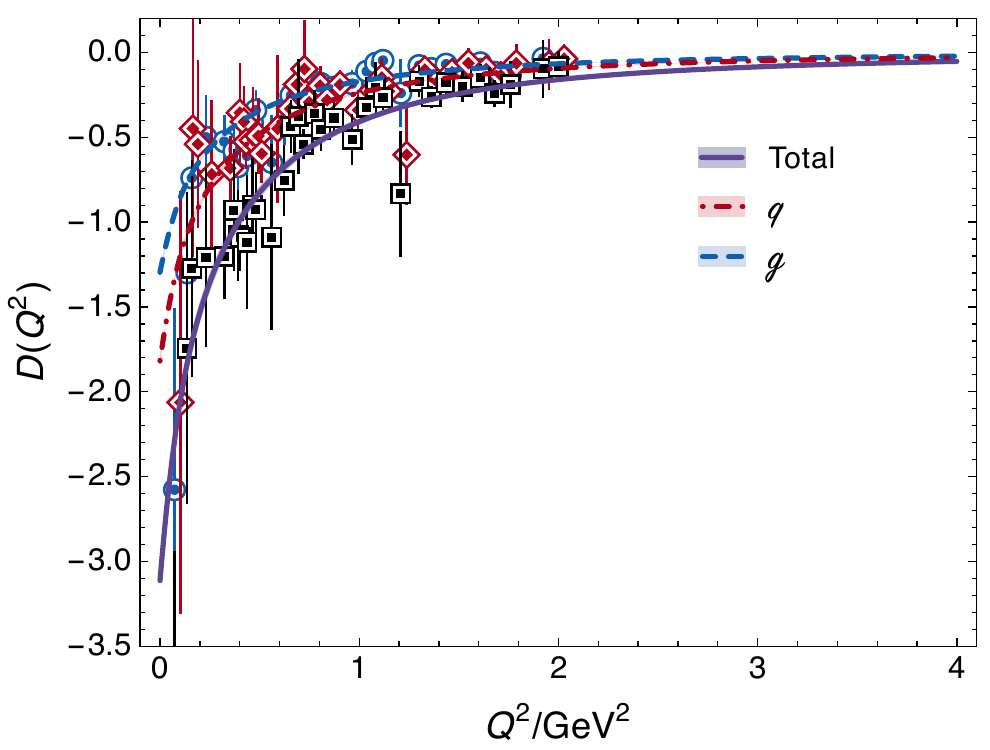} 
    \end{tabular}
    \caption{Nucleon GFFs. Scale-invariant (total) GFF, and quark and gluon decomposition at $\zeta=2$ GeV.}
    \label{fig:NucleonGFFs}
\end{figure}

\section{Conclusions}
We described a unified set of CSM predictions for $\pi,K$ and nucleon GFFs. 
The results were obtained using a symmetry-preserving formulation of all quantum field equations relevant to the analysis. 
For the systems under scrutiny, as inferred from the corresponding form factors, the charge distributions extend over a larger region than the mass distributions. 
Furthermore, as revealed by pressure densities, shear forces are maximal where confinement forces become dominant and, notably, near-core pressures are significantly greater than those expected in neutron stars. 
Naturally, kaon profiles are more compact than those of the pion. 
For the nucleon, CSMs predict $-D^\textbf{N}(0)=3.114(10)$. 
All produced GFFs satisfy expected symmetry requirements.

\section*{Acknowledgments}
Work supported by: 
National Natural Science Foundation of China (grant no.\ 12135007, 12233002);
Natural Science Foundation of Jiangsu Province (grant no.\ BK20220122); Helmholtz-Zentrum Dresden-Rossendorf, under the High Potential Programme;
Spanish Ministry of Science and Innovation (MICINN grant no.\ PID2022-140440NB-C22); and Junta de Andalucía (grant no.\ P18-FR-5057).

\bibliographystyle{unsrt}
\bibliography{bibliography}

\end{document}